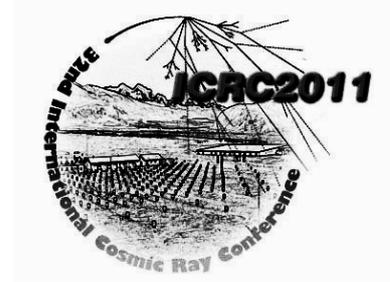

# Design and Fabrication of Detector Module for UFFO Burst Alert & Trigger Telescope

A. JUNG[1], S. AHMAD[2], K.-B. AHN[3], P. BARRILLON[2], S. BLIN-BONDIL[2], S. BRANDT[4], C. BUDTZ-JØRGENSEN[4], A.J. CASTRO-TIRADO[5], P. CHEN[6], H.S. CHOI[7], Y.J. CHOI[8], P. CONNELL[9], S. DAGORET-CAMPAGNE[2], C. DE LA TAILLE[2], C. EYLES[9], B. GROSSAN[10], I. HERMANN[8], M.-H. A. HUANG[11], S. JEONG[1], J.E. KIM[1], S.-W. KIM[3], Y.W. KIM[1], J. LEE[1], H. LIM[1], E.V. LINDER[1,10], T.-C. LIU[6], N. LUND[4], K.W. MIN[8], G.W. NA[1], J.W. NAM[1], K.H. NAM[1], M.I. PANASYUK[12], I.H. PARK[1], V. REGLERO[9], J.M. RODRIGO[9], G.F. SMOOT[1,10], Y.D. SUH[8], S. SVERTILOV[12], N. VEDENKIN[12], M.-Z WANG[6], I. YASHIN[12], M.H. ZHAO[1], FOR THE UFFO COLLABORATION.

[1]*Ewha Womans University, Seoul, Korea*
[2]*University of Paris-Sud 11, France*
[3]*Yonsei University, Seoul, Korea*
[4]*National Space Institute, Denmark*
[5]*Instituto de Astrofisica de Andalucia, Consejo Superior de Investigaciones Cientificas (CSIC), Spain*
[6]*National Taiwan University, Taipei, Taiwan*
[7]*Korea Institute of Industrial Technology, Ansan, Korea*
[8]*Korea Advanced Institute of Science and Technology, Daejeon, Korea*
[9]*University of Valencia, Spain*
[10]*University of California, Berkeley, USA*
[11]*National United University, Miao-Li, Taiwan*
[12]*Moscow State University, Moscow, Russia*
*erajung@gmail.com*

**Abstract:** The Ultra-Fast Flash Observatory (UFFO) pathfinder is a space mission devoted to the measurement of Gamma-Ray Bursts (GRBs), especially their early light curves which will give crucial information on the progenitor stars and central engines of the GRBs. It consists of two instruments: the UFFO Burst Alert & Trigger telescope (UBAT) for the detection of GRB locations and the Slewing Mirror Telescope (SMT) for the UV/optical afterglow observations, upon triggering by UBAT. The UBAT employs a coded-mask γ/X-ray camera with a wide field of view (FOV), and is comprised of three parts: a coded mask, a hopper, and a detector module (DM). The UBAT DM consists of a LYSO scintillator crystal array, multi-anode photo multipliers, and analog and digital readout electronics. We present here the design and fabrication of the UBAT DM, as well as its preliminary test results.

**Keywords:** Ultra-Fast Flash Observatory (UFFO) pathfinder, Gamma-Ray Bursts (GRBs), UFFO Burst Alert & Trigger telescope (UBAT), Slewing Mirror Telescope (SMT), coded-mask, LYSO, multi-anode photo multipliers, detector module (DM)

## 1 Introduction

Gamma-Ray Bursts (GRBs) are by far the most energetic events in the sky, with higher peak photon luminosities than any other objects in the universe. They also possess a wide range of redshift factors, from less than one to above eight. Hence, GRB physics exists at the very ends of extreme physics, in terms of both energy and time in the history of our universe.

GRBs have been classified into two types according to their γ/X-ray emission characteristics: a short burst with a hard spectrum, typically with a duration of less than two seconds, and a longer, softer type burst [1]. Nevertheless, it is not well known what makes them different, and thus the origin and evolution of GRBs are still big mysteries in high-energy astrophysics.

As the early-stage observations of the light curves and emission spectrum are essential to the understanding of the nature of GRBs, there have been significant efforts for coordination of satellite and ground based observations. However, the early-phase observations were often limited by the response of the instruments operated in space. In the case of *Swift* observatory, for example, the entire spacecraft has to be maneuvered to point toward the GRB position after being triggered by the detection of γ-ray bursts, which takes about a minute. Thus, it is difficult to obtain the sub-minute ultraviolet (UV) and visible light curves with *Swift* [1]. In this regard, it should be noted that the proposed Ultra-Fast Flash Observatory (UFFO) pathfinder is designed to observe sub-minute GRB light curves by adopting the slewing mirror technology.

The UFFO pathfinder is planned to be launched by the *Lomonosov* spacecraft in November 2011 into a sun-synchronous orbit at an altitude of ~550km. The UFFO pathfinder consists of two instruments: the UFFO Burst Alert & Trigger telescope (UBAT) for the detection of GRB locations, and the Slewing Mirror Telescope (SMT) for the UV/optical afterglow observations [2, 3], upon triggering by UBAT.

The UBAT detects the γ/X-ray photons from GRBs in the energy range of ~5 to ~200 keV (or ~100 keV with employment of an electrical veto system to reduce the effect of internal backgrounds). The LYSO scintillation crystal converts the γ/X-ray photons to UV photons, which eventually become electrical pulses through the chain of multi-anode photomultipliers and pulse-shaping electronics. These electrical pulses are measured and recorded with a period of 2.5 μs.

The UBAT is comprised of a mechanical assembly of the coded mask with a hopper and a detector module (DM). The coded mask, located on top of the hopper, is an array of opaque and transparent elements and gives the information on the location of the GRB source by projecting the γ/X-ray images onto the detector plane. The hopper, situated between the coded mask and the UBAT DM structure, supports the coded mask mechanically and acts as a collimator.

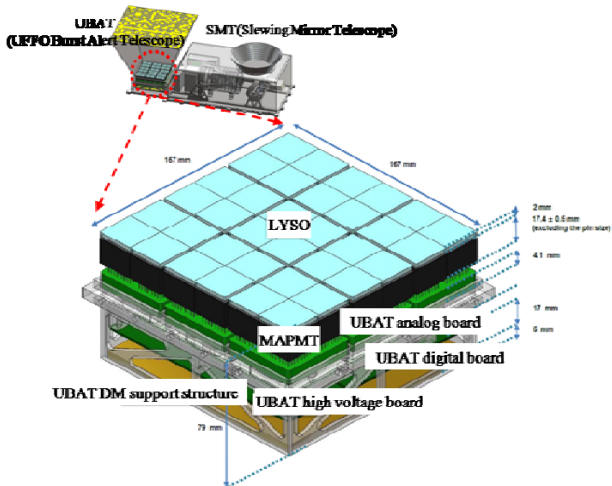

Figure 1. UFFO pathfinder and UBAT DM

## 2 UBAT Detector Module (DM)

The UBAT DM includes the LYSO crystal, MAPMTs, analog and digital boards, high voltage boards, and the support structure, as can be seen in Figure 1.

When the γ/X-ray photons hit the LYSO crystal, it produces UV photons by scintillation in proportion to the energy of the incident γ/X-ray photons. These UV photons are converted into electrons, which are multiplied by the MAPMTs and fed into the analog boards. The analog signals are then processed at the digital boards, where the trigger algorithm is also performed.

### 2.1 Specifications

The specifications of the UBAT DM are shown in Table 1.

| Field of View | ~ 1.8 sterad (90.2° x 90.2°) |
|---|---|
| Point Spread Function | ≤ 10 arcmin |
| Detector | LYSO + MAPMT |
| Detector energy range | 5 - 200 keV |
| Number of pixels | 48 x 48 |
| Pixel size | 2.88 x 2.88 x 2 mm$^3$ |
| Effective area | 191.1 cm$^2$ |
| Energy resolution | 2 keV (FWHM) at 60keV |
| Quantum efficiency | 99 % at 100keV |
| Sensitivity | 310 mCrab with 10s exposure at 5.5σ for 4 ~ 50keV |
| UBAT total mass | 10 kg |
| UBAT DM mass | 2.6 kg |
| Volume | 400.8 x 400.8 x 365 mm$^3$ |
| Power consumption | 10 W |

Table 1. Specifications of UBAT DM

### 2.2 LYSO

Table 2 shows the physical and scintillation properties of LYSO.

| Density | 7.2 g/cm$^3$ | Effective Z number | 66 |
|---|---|---|---|
| Peak wavelength | 420 nm | Decay constant | 42 ns |
| Reflective index (at 420nm) | 1.82 | Light yield (photons/MeV) | ~ 2.5x10$^4$ |
| Melting point | 2,100 °C | Hygroscopicity | No |
| Light output (compared with NaI(Tl)) | 73 ~ 75% | | |

Table 2. Properties of LYSO

LYSO has many advantages over other gamma spectrometers. We have listed some of them as follows:

- high stopping power due to its high density
- no need for a cooling system
- high spatial resolution by adopting pixelized crystals
- simultaneous measurement of X-rays and γ- rays
- high light yield
- fast decay time

LYSO has internal background coming from the beta decay of $^{176}$Lu, with ~300 counts/sec/cm$^3$. However,

most of the background is in the high energy regime above 100 keV. If the energy is limited to below 100 keV, the internal background rate drops to ~5 counts/sec/cm$^3$, which is similar to the astronomical γ/X-ray background below 100 keV. We employ electrical veto methods to limit our photon detection below 100 keV.

### 2.3 Multi-anode photomultiplier tubes (MAPMTs)

The UBAT DM consists of 36 modules of MAPMTs, with each module composed of 8 x 8 channels. The 36 MAPMT modules are arranged in a square with 6 rows and 6 columns, hence the total 2304 MAPMT channels makes an array of 48 x 48 pixels. Located underneath the LYSO crystal array, each MAPMT channel corresponds to a single LYSO crystal element. The photocathode is made of a bialkali material and the number of dynode stages is 12. Four MAPMT packages are connected to a single UBAT analog board. Table 3 shows the specifications of the MAPMT.

| Spectral response range | | 185 ~ 650 nm |
|---|---|---|
| Window material/Thickness | | Ultra violet glass/0.8 mm |
| photocathode | Material | Bialkali |
| | Minimum effective area | 23 x 23 mm$^2$ |
| Dynode structure | | Metal channel dynode |
| Weight | | 27 g |
| Number of dynode stages | | 12 |
| Operating ambient temperature | | -30 ~ 50 ℃ |
| Storage temperature | | -30 ~ 50 ℃ |
| Supply voltage between anode and cathode (V/gain) | | 1,100/1 x 10$^6$ |
| Average anode current | | 0.1mA |

Table 3. Specifications of MAPMT

### 2.4 UBAT analog board

The purposes of the analog boards are photon counting and measurement of photon energies. The photoelectron signals from the MAPMTs are converted into digital signals using application-specific integrated circuits, which are measured with an interval of 2.5 μs. The total charges from the MAPMT are summed for each photon event, which is proportional to the incident photon energy and is converted to a pulse duration time (Q-to-T conversion) by the process called KI [4]. The signals from the MAPMT anodes are fed into the ASIC preamplifiers of the analog boards for signal processing, with adjustable gain control to compensate for the gain non-uniformity of the MAMPTs. There are nine UBAT analog boards, with each board handling four MAPMTs with four ASIC preamplifiers.

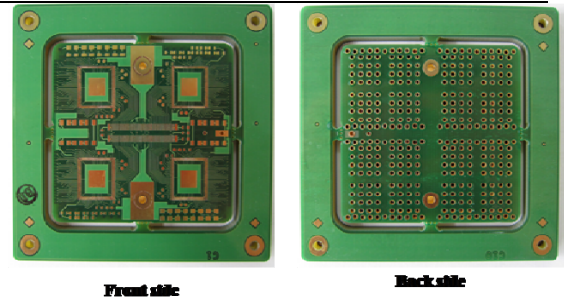

Figure 2. Fabricated UBAT analog board PCBs

### 2.5 UBAT digital board

There is only one UBAT digital board which handles all the UBAT analog boards. The hardware of the UBAT digital board electronics includes FPGA (Field Programmable Gate Array) chips, which perform the trigger algorithm, and the interface with the UBAT analog boards. The trigger algorithm proceeds as follows: when GRBs are detected, a trigger flag is generated and image reconstruction starts by performing cross-correlations between the shadowgram and the mask patterns. Housekeeping and other operation-related functions are also controlled by the FPGA chips. The physical layout of the UBAT digital board is shown in Figure 3.

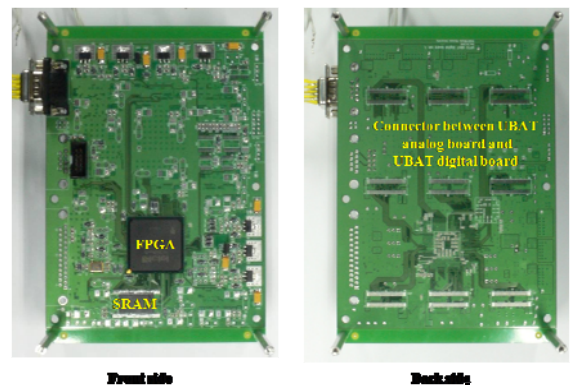

Figure 3. Prototype UBAT digital boards

### 2.6 UBAT high voltage board

The UBAT high voltage board provides high voltages to the MAPMT dynodes, which operate at the nominal voltage of ~900 V. EMCO chips are employed to raise the voltage from 10 V, which is supplied from the power distribution board of the UFFO Data Acquisition system. A Gain Control Logic is used to adjust the high voltages.

### 2.7 UBAT DM support structure

The UBAT DM is integrated into the UBAT DM support structure, as sketched in Figure 4, which not only allows tight packing to endure vibrations at the time of launch but also provides a path to efficiently evacuate the heat generated from the internal circuits. Figure 5 shows an assembly procedure and the assembled structure: a coded mask, hopper, UBAT mechanical box and UBAT DM.

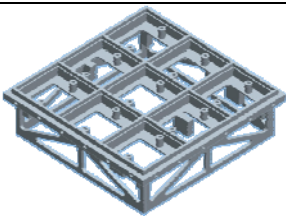

Figure 4. Sketch of the UBAT DM support structure

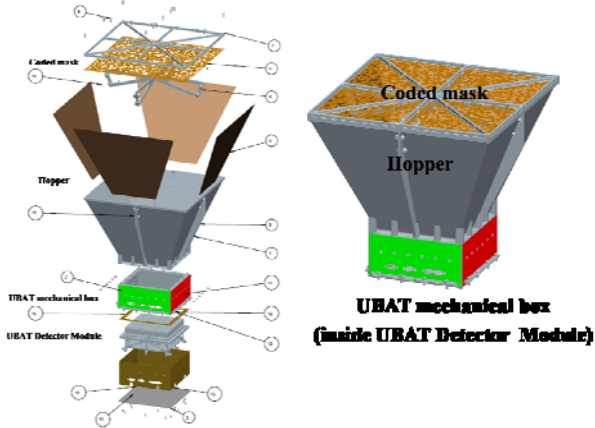

Figure 5. Assembly procedure of UBAT

## 3  UBAT interface test

We have carried out interface tests between the UBAT analog boards and the UBAT digital board. The objectives of the interface tests are:
- to verify the interface between the UBAT analog board and the UBAT digital board
- to confirm the performance of the analog board in response to the light signal

The test setup is shown in Figure 6. We used an LED light source for an input signal, and USB 8451(NI) for the control and data transfer to PC. A light signal from the LED is reflected by a mirror and hit the pixels of the MAPMTs. The illuminated regions on the MAPMT surface are changed by rotating the mirror.

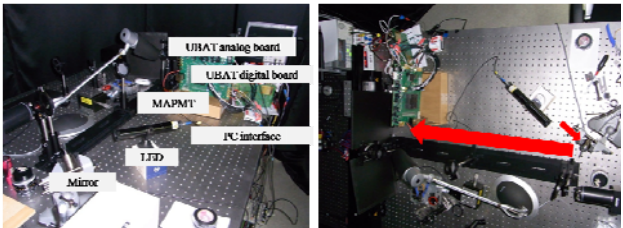

Figure 6. Setup for the interface tests

Figure 7 shows the test results: the LED light images move downwards in the top panels (a-1 through a-3) and from the lower left corner to the upper right corner in the bottom panels (b-1 through b-3), in accordance with the changes of the illuminated region on the MAPMT surface.

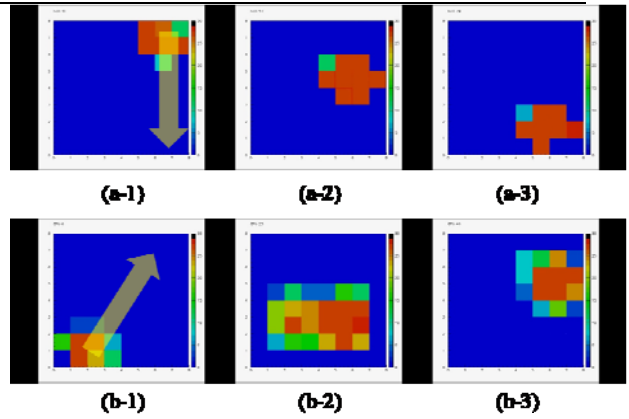

Figure 7. Images seen on the PC screen according to the changes of the illuminated region on the MAPMT

## 4  Conclusions

We developed a γ/X-ray burst telescope, which is called the UFFO pathfinder, that can observe the early phase of short duration gamma ray bursts with a sub-minute response time. The fast detection mechanism is realized by employing the slewing mirror technique. The UFFO pathfinder consists of two instruments: the UFFO Burst Alert & Trigger telescope (UBAT) for the detection of GRB locations and the Slewing Mirror Telescope (SMT) for the UV/optical afterglow observations. The UBAT is comprised of three parts: a coded mask, a hopper, and a detector module (DM). We have described the subcomponents of the UBAT DM in this report: a LYSO scintillator crystal array, multi-anode photomultipliers, and analog and digital readout electronics. We are in the final stages of development and expect to deliver the instrument for integration with the spacecraft in two months.

References

[1] I.H. Park et al., THE UFFO (ULTRA-FAST FLASH OBSERVATORY) PATHFINDER, arXiv:0912.0773
[2] S. Jeong et al., Optical Performances of Slewing Mirror Telescope for UFFO-Pathfinder, Proceedings of this Conference, # 1269
[3] J.E. Kim et al., Implementation of the readout system in the UFFO Slewing Mirror Telescope, Proceedings of this Conference, # 1263
[4] S. Ahmad et al., SPACIROC: A Rad-Hard Front-End Readout chip for the JEM-EUSO telescope, 2010 JINST 5 C12012